\documentstyle[aps,preprint]{revtex} 
\begin{document}

\def\half{{1 \over 2}}

\def\nfdf{${\cal N}=4, d=4$}
\def\ymt{Yang-Mills theory}
\def\sun{SU($N$)}
\def\laa{\langle\kern-.3em \langle}
\def\raa{\rangle\kern-.3em \rangle}
\def\la{\lambda}
\def\cn{{\cal N}}\def\cz{{\cal Z}}
\def\ymt{Yang-Mills theory}
\def\sftH{string field theory Hamiltonian}
\def\aaa#1{a^{\dagger}_{#1}}
\def\aap#1#2{(a^{\dagger}_{#1})^{#2}}
\def\ee{\hbox{e}}
\def\dd{\hbox{d}}\def\DD{\hbox{D}}
\def\tr{\hbox{tr}}
\def\Tr{\hbox{Tr}}
\def\part{\partial}
\def\dj#1{{\delta\over{\delta J^{#1}}}}
\def\cO{{\cal O}}
\def\ssc{\scriptscriptstyle}

\title{D-brane charges and $K$-homology}
 
\author{Vipul Periwal}

\address{Department of Physics, 
Joseph Henry Laboratories,
Princeton University, 
Princeton, NJ 08544 
\tt vipul@princeton.edu}

\maketitle 
\begin{abstract}
It is argued that D-brane charge takes values in $K$-homology.  For 
smooth manifolds with spin structure, this could explain why the
phase factor $\Omega(x)$  calculated with a D-brane state $x$ in IIB theory 
appears in Diaconescu, Moore and Witten's computation of 
the partition function of IIA string theory. 
\end{abstract}
\bigskip
\tightenlines
\def\st#1#2{#1\star#2}
\def\poi#1#2{\{#1,#2\}}
\def\cf{Cattaneo-Felder }
\def\sw{Seiberg-Witten }

Diaconescu, Moore and Witten\cite{dmw}\ recently compared the partition 
functions of IIA string theory and $M$-theory.  They found agreement, 
amusingly interpreted as a derivation of $K$-theory from $M$-theory.
Many subtle mathematical effects have to conspire for this agreement, 
so it seems important to explore further the building blocks of their
results.

$K$-theory will enter into the discussion in a variety of ways so
here is a summary of $K$-theory in string theory.  Witten\cite{wk}\ 
argued that D-brane charge should take values in $K$-theory, 
following earlier work by Cheung and Yin\cite{cy}\ and Moore and 
Minasian\cite{mm}, and providing a mathematical setting  
for the results of Sen\cite{sen}.  To be precise, the groups 
$K^{0(1)}(X)$ are associated with  
D-branes in IIB(A) string theory on the spacetime $X,$ respectively.  
In later work, Moore and 
Witten\cite{mw}\ argued that Ramond-Ramond fields should take values in
$K$-theory as well, with $K^{1(0)}(X)$ classifying these fields in 
IIB(A) string theory.

An important point in 
\cite{dmw}\ is the computation of a phase factor $\Omega(x)$ 
associated with an element $x$ which is a Ramond-Ramond field in
IIA string theory on a manifold $X.$    Somewhat surprisingly,
this phase factor is computed from a D-brane state in IIB string
theory, also associated with $x$ by means of the above arguments.  
The perplexing appearance of IIB string theory in a computation that {\it a 
priori} involves only IIA theory is pointed out in footnote 14 in 
\cite{dmw}.

Regressing a little bit to Polchinski's basic operational definition of 
D-branes\cite{p},  recall that string theory in the presence of a D-brane
is  na\"\i vely defined by specifying a submanifold $M$ of $X$ and 
including open strings which have both ends on $M.$   
Suppose we act with a diffeomorphism 
\begin{equation}
\phi:X \rightarrow X',
\end{equation}
then the submanifold 
\begin{equation}
M\mapsto  \phi_{*}(M)\subset X'
\end{equation}
and this map is covariant, equivalently $M$ is `pushed-forward' by 
$\phi_{*}.$  On the other hand, if there are forms  $G$ defined on $X'$ 
these fields map contravariantly to define `pulled-back' forms:
\begin{equation}
\phi^{*}G(y)(v_{1},\ldots,v_{n}) \equiv\ G(\phi(y))(\phi_{*}v_{1},\ldots,
\phi_{*}v_{n}) 
\end{equation}
for $v_{i}(y)$ tangent vectors at $y$ in $X,$ and $\phi_{*}v_{i}$ the 
corresponding pushed-forward vectors.

Now, $K$-theory is an extraordinary cohomology theory for manifolds, 
and as such transforms contravariantly, {\it i.e.}
$\phi$ induces a map
\begin{equation}
\phi^{*}: K(X') \rightarrow K(X).
\end{equation}
This is perfectly appropriate for Ramond-Ramond 
fields, but for making contact between the covariant operational definition of 
D-branes and contravariant $K$-theory one needs additional structure such as
Poincar\'e duality, as in section 5.1 of \cite{dmw}\ for example.  
Such additional structure is always available  
when one works on $\hbox{spin}_{c}$ manifolds for example, so this is 
not a major assumption in a physical context.  However, it is 
important to keep this assumption in mind since it implies that the
correspondence between D-branes, as characterized by Polchinski, and
D-brane charges, if associated with elements of $K(X),$\cite{wk}\ 
is not canonical.

It is the purpose of this note to explain that Poincar\'e 
duality in $K$-theory makes it more natural for D-brane charges to 
take values in $K$-homology groups, where $K$-homology is the 
homology theory dual to $K$-theory.  This is to be contrasted to
using Poincar\'e duality in homology-cohomology and then lifting to
$K$-theory.  The utility 
of this point of view is that it will then become apparent why a D-brane 
{\it state in IIB string theory} is necessary for defining a phase factor 
associated with a {\it Ramond-Ramond field in IIA theory}.  Looking a little
further, it is to be hoped that computing in an appropriately defined
$K$-homology theory for conformal field theories might lead to some 
insight into possible D-brane states in an abstract closed string 
background, perhaps explaining the Cardy conditions\cite{c}\ 
as characterizing a cycle in $K$-homology. 

A note on convention: upper indices are contravariant and lower ones 
covariant.  In particular, we have  $K_{0}(C(X)) = K^{0}(X),$ where 
$C(X)$ is the algebra of continuous functions on the manifold $X.$

The basic idea for defining $K$-homology groups is 
the following\cite{a,bdf,kas}:  For any elliptic operator $D$ on
a manifold $X,$ and a smooth vector bundle $E$ over $X,$ one can define 
an operator $D_{E}$ acting on sections of $E$ (by using partitions of 
unity or a connection on $E,$ for example).  
The operator $D_{E}$ depends on the choices but it is a Fredholm 
operator ({\it i.e.} has a finite-dimesional kernel and cokernel).  
Its index does not depend on these choices, so we get a map
\begin{equation}
{\rm Index}_{D}: K^{0}(X) \rightarrow {\bf Z} 
\end{equation}
with ${\rm Index}_{D}(E) \equiv {\rm Index}(D_{E}).$
It is crucial in this construction that the operator $D$ is a 
pseudolocal operator, in other words it commutes with the operation of 
multiplication by functions on $X$ up to compact operators.  
Stated algebraically, 
$K^{0}(X)$ can
be interpreted as equivalence classes of projection operators $p$ on a
Hilbert space, and elements of $K_{0}(X)$ are suitably defined 
equivalence classes of abstract pseudolocal Fredholm operators $F$ on $X.$  
The natural pairing between these two sets of objects is given by the
index map by computing ${\rm Index}(pFp).$ 
In topological $K$-theory expressed in terms of algebras of functions, 
the definition of $K$-homology involves 
classifying extensions of $C(X)$ by the algebra of compact operators 
up to unitary equivalence\cite{bdf}.  

This example serves as the intuition for Kasparov's definition of
$K\!K$-groups, which seem to be the most natural framework for our 
purposes.
The precise definitions  are a little 
technical\cite{higson,bla,con,jt}\ so all mathematical precision has 
been eliminated from the following discussion.  The tensor products 
are all ${\bf Z}_{2}$ graded tensor products. 
$K\!K^{*}(A,B)$ is an Abelian group depending contravariantly on the 
algebra $A$ and covariantly on the algebra $B.$  The elements of $K\!K^{*}(A,B)$
are homotopy classes of generalized elliptic operators over $A$
with coefficients in $B$ given by (i) an operator $F$ between
two $B$-modules ${\cal E}_{0,1}$ and representations of $A$ as 
operators on ${\cal E}_{0,1}$ such that $[a,F], a(F^{*}F-1)$ and
$a(FF^{*}-1)$ are all compact operators.  The condition $[a,F]$ 
compact is obviously a generalization of the pseudolocal property in 
the example above, and the other two conditions require $F$ to be 
unitary up to compact operators.  The facts of interest 
to us are the following:
\begin{eqnarray}
\hbox{For}\ A={\bf C},\ \ K\!K^{*}({\bf C}, B) = K_{*}(B).\\
\hbox{For}\ B={\bf C},\ \ K\!K^{*}(A, {\bf C}) = K^{*}(A). 
\end{eqnarray}
In addition, there is a bilinear associative intersection product
\begin{equation}
K\!K^{n}(A_{1},B_{1}\otimes D)\otimes_{D} K\!K^{m}(D\otimes A_{2},B_{2}) 
\rightarrow 
K\!K^{{n+m}}(A_{1}\otimes A_{2},B_{1}\otimes B_{2}),
\end{equation}
with addition  mod 2 in the superscript.  $K\!K$-equivalence of $A$ 
and $B$ is the statement that the $K\!K$ groups of $A\otimes E$ and 
$B\otimes E$ with any
other algebras $D,E$ are isomorphic with $A\otimes E$ or $B\otimes E$ 
as contravariant (resp. 
covariant) arguments and $D$ as the covariant (resp. contravariant) 
argument.  Finally, $K\!K(A,A)$ is a ring with unit.

Suppose now that we have two algebras $A$ and $B$ such 
that there are elements 
\begin{equation}
\alpha\in K\!K(A\otimes B,{\bf C}),\quad \beta\in K\!K({\bf C},A\otimes B)
\end{equation}
with the property that 
\begin{equation}
\beta\otimes_{A}\alpha = 1_{B} \in\ K\!K(B,B), \\
\beta\otimes_{B}\alpha = 1_{A} \in\ K\!K(A,A).
\end{equation}
We then get $K\!K$-duality isomorphisms 
\begin{equation}
K_{*}(A) \cong K^{*}(B) ,\ \hbox{and} \ 
K^{*}(A) \cong K_{*}(B)
\label{duality}
\end{equation}
between the $K$-theory($K$-homology) of $A$ and the 
$K$-homology($K$-theory) of $B.$  The algebras $A$ and $B$ are
Poincar\'e dual\cite{con}. 
In general these algebras are {\it not} $K\!K$-equivalent. 

An example of this duality can be constructed as 
follows\cite{kas,con,bla}: Suppose 
that $E$ is the total space of a real vector bundle with a fibre metric 
over a differentiable manifold $X.$  
Consider the algebra of the sections of the
complex Clifford bundle $A\equiv\Gamma({\rm Cliff}(E))$  
and the algebra $B\equiv C(X).$  
Using a partition of unity subordinate to an open cover of $X$ such 
that $E$ is trivializable over the sets in the open cover, we can 
define an operator $d$ which restricts  to the exterior derivative in 
the fibre direction locally.  $d$ acts on the Hilbert space of 
square integrable sections of the complexification of $\bigwedge^{*}E.$
Furthermore, $Y\equiv \Gamma({\rm Cliff}(E))\otimes C(X)$ acts by
Clifford multiplication on these sections.  
We set $F=(d+\delta)/(1+(d+\delta)^{2})^{1/2},$ where $\delta$ is the 
adjoint of $d.$  This defines $\alpha\in K\!K(Y,{\bf C}).$
For $\beta$ we note that the algebra of functions on $E$ that vanish at infinity
$C_{0}(E)$ is $K\!K$-equivalent to
$A,$ so we take the Hilbert module to be $Y$ itself, and construct an
element in $K\!K({\bf C},C_{0}(E)\otimes B) \cong K\!K({\bf C},Y)$ by
taking $F$ to be multiplication by a parametrized  version of the Bott element 
\begin{equation}
F = v_{p}(1+|v_{p}|_{p}^{2})^{-1/2}
\end{equation}
for $p$ a point in $X,$  $v_{p}$ a vector in $E_{p},$ with the norm 
computed with the fibre metric at $p.$ 
 
If $E$ has a $\hbox{spin}_{c}$ structure, $\Gamma({\rm Cliff}(E))$
is $K\!K$-equivalent to
$C(X)$ when $E$ has even-dimensional fibres, and $K\!K$-equivalent to
$C(X)\hat\otimes {\bf C}_{1}$ when $E$ has odd-dimensional 
fibres\cite{bla}, 
where ${\bf C}_{1}$ is the Clifford algebra for ${\bf C},$  the 
superalgebra with one generator of odd degree $\sigma:\sigma^{2}=1.$ 
Since $K\!K$-groups have the appropriate Clifford periodicity
properties, these algebras are $K\!K$-equivalent up to the shift
for the odd-dimensional case.

For the application to type II string theory, a fairly general
context in which we expect IIA and IIB duality is one in which there 
is a fixed-point free action of ${\rm U}(1)^{2n+1}.$  This defines a 
sub-bundle $E$ of $TX$ with fibres of dimension $2n+1.$  Then it 
is natural to conjecture that at the level of $K\!K$-theory, the 
$K\!K$-duality with the shift due to the odd dimension of the bundle gives 
the expected IIA-IIB duality.  
For strings in ten dimensions, this 
gives an isomorphism between  Ramond-Ramond fields in IIA string 
theory and  D-brane charges in IIB string theory as desired\cite{dmw}.
I want to emphasize that I have {\it not} shown that
this is in fact what we mean by IIA-IIB duality in string theory---I 
am merely pointing out that this would make sense of the 
computation in \cite{dmw}, as well as identifying
D-brane charge in a way that is consistent with the covariance of the physical 
description of D-branes\cite{p}.
Of course, a U(1) action is not general enough to encompass all 
manifestations of IIA-IIB duality.  The general setting is probably 
that of Takai duality\cite{bla}\ for ${\bf Z}_{2}$ but with specific
constraints so that the Takai dual is also the $K\!K$-dual.  

To conclude,  just considering 
Ramond-Ramond fields  classified by   $K\!K({\bf C},X),$ or just
D-branes  classified by $K\!K(X,{\bf C}),$ independently may be inadequate for 
physics.  As an example, it seems to me  it should be possible
to describe the jump 
in Ramond-Ramond fields across a D-brane either in terms of 
the local image in cohomology of two different $K$-theory 
classes (see \cite{mw} for example), or 
by considering an appropriate element of $K\!K(X,X).$  This ring has a 
natural action on both $K^{0}(X)$ and $K_{0}(X)$ via the Kasparov 
product, and it would be interesting to figure out if this
action is of physical significance.  More concretely, understanding the 
Cardy conditions defining consistent boundary states as the data
defining $K$-cycles seems worth pursuing.  

Acknowledgements: 
I am grateful to M. Berkooz for helpful conversations.
This work was supported in part by NSF grant PHY98-02484.

\def\np#1#2#3{Nucl. Phys. B#1,  #3 (#2)}
\def\prd#1#2#3{Phys. Rev. D#1, #3 (#2)}
\def\prl#1#2#3{Phys. Rev. Lett. #1, #3 (#2)}
\def\pl#1#2#3{Phys. Lett. B#1, #3 (#2)}
\def\jhep#1#2#3{J. High Energy Phys. #1, #3 (#2)}

\end{document}